\documentclass[12pt]{article}
\usepackage{graphicx}
\usepackage{lineno}

\newcommand\pubnumber{}
\newcommand\pubdate{\today}

\def\institute{Laborat\'{o}rio de Instrumenta\c{c}\~{a}o e F\'{i}sica Experimental de Part\'{i}culas, Braga, Portugal}
\def\support{\footnote{The author is funded by the grant LIP/50007 through FCT, COMPETE2020-Portugal2020, FEDER, POCI-01-0145-FEDER-007334 and by the FCT project CERN/FIS-NUC/0005/2015.}}

\def\Title#1{\begin{center} {\Large #1 } \end{center}}
\def\Author#1{\begin{center}{ \sc #1} \end{center}}
\def\Address#1{\begin{center}{ \it #1} \end{center}}

\newcommand\pubblock{\rightline{\begin{tabular}{l} \pubnumber\\
         \pubdate  \end{tabular}}}
\newenvironment{Abstract}{\begin{quotation}  }{\end{quotation}}
\newenvironment{Presented}{\begin{quotation} \begin{center} 
             PRESENTED AT\end{center}\bigskip 
      \begin{center}\begin{large}}{\end{large}\end{center} \end{quotation}}

\def\beq{\begin{equation}}
\def\eeq#1{\label{#1}\end{equation}}
\def\eeqn{\end{equation}}

\def\beqa{\begin{eqnarray}}
\def\eeqa#1{\label{#1}\end{eqnarray}}
\def\eeqan{\end{eqnarray}}

\let\bar=\overbar

\def\Dslash{\not{\hbox{\kern-4pt $D$}}}
\def\dslash{\not{\hbox{\kern-2pt $\del$}}}

\def\msb{{\bar{\ssstyle M \kern -1pt S}}}

\def\met{E_{\rm miss}^{\rm T}}
\def\pt{p_{\rm T}}

\begin{document}
\begin{titlepage}
\pubblock

\vfill
\Title{Searches for flavour changing neutral currents in the top sector}
\vfill
\Author{ J. P. Araque\support\\On behalf of the ATLAS and CMS collaborations}
\Address{\institute}
\vfill
\begin{Abstract}
	Flavour Changing Neutral Current (FCNC) processes are forbidden at tree level in the Standard Model and highly suppressed at higher orders. This makes FCNC one of the key processes to search for new physics since any small deviations from the Standard Model expectations could have a big impact. Both ATLAS and CMS Collaborations  have designed a comprehensive strategy to search for FCNC in top quark physics both in the production and decay. The strategies followed by both collaborations are here described, using data from $pp$ collisions at the LHC collected at a centre of mass energies of 7 and 8~TeV with integrated luminosities ranging from $5~\rm{ fb}^{-1}$ to $20.3~\rm{ fb}^{-1}$.
	\end{Abstract}
\vfill
\begin{Presented}
$9^{th}$ International Workshop on Top Quark Physics\\
Olomouc, Czech Republic,  September 19--23, 2016
\end{Presented}
\vfill
\end{titlepage}
\def\thefootnote{\fnsymbol{footnote}}
\setcounter{footnote}{0}

\section{Introduction}

In the Standard Model (SM) FCNC processes are forbidden at tree level by the GIM mechanism and highly suppressed at higher orders. This implies that expected cross sections for such processes are very small when only considering SM interactions. Nonetheless different models beyond the SM such as the MSSM, models with 2 Higgs doublets or models including extensions of the lepton and quark sectors, predict an increase of several orders of magnitude above the SM expectations, proving a powerful test for new physics phenomena. Both the ATLAS~\cite{ATLAS} and CMS~\cite{CMS} collaborations have defined comprehensive search strategies for FCNC processes both in the production and decay of the top quark, covering all possible channels associated with FCNC processes ($Z$, $H$, $\gamma$ and $g$ channels). 

\section{FCNC processes in the $Z$ boson channel}

The ATLAS Collaboration~\cite{ATLASZ} published a search for FCNC processes in the production of a $t\bar t$ pair in data collected at $\sqrt{s}=8$~TeV with an integrated luminosity of $20.3~\rm{fb}^{-1}$. One of the top quarks decays through the usual SM channel ($t\rightarrow bW$) while the other top quark will decay through an FCNC process to $Zq$. This setup presents a tri-leptonic topology (both $W$ and $Z$ are required to decay leptonically) with high missing transverse energy ($\met > 20$~GeV). Additional requirements are made in multiplicity of jets (at least 2) and jets tagged as originating from a $b$-quark ($b$-tagged jets). The event is reconstructed by minimising a $\chi^2$ which contains information from the different reconstructed objects (both top quarks and the $W$ boson) considering all possible combinations of leptons and jets. Observed upper limits for the branching-ratio (BR) of $t\rightarrow qZ$ at a 95\% confidence level (C.L.) are set at $7\times10^{-4}$.




The CMS Collaboration~\cite{CMSZ} presents a search for FCNC processes via the $Z$ boson both in the decay of the top quark, with a $t\bar t$ channel similar to the one previously described, with the addition of a single-top channel in which the FCNC process takes place in production. Data collected at $\sqrt{s}=8$~TeV and with an integrated luminosity of $19.7~\rm{fb}^{-1}$ were analysed. Both channels focus on a tri-lepton topology, requiring leptonic decays of the $Z$ boson and the top quark with high missing energy ($\met > 40$~GeV). Jets and $b$-tagged jets multiplicities are used as well in both channels. Two boosted-decision trees (BDT) are defined using kinematic variables from the $Z$ boson, top quark and jets and $b$-tagged jets to better differentiate signal from background. The output of the BDTs are used as discriminant variables and observed upper limits are set, at a 95\% C.L., for BRs above $2.2\times10^{-4}$ ($4.9\times10^{-4}$) for the $t\rightarrow Zu$ ($t\rightarrow Zc$) channel. Figure~\ref{fig}~(top-left) shows the the excluded region as a function of the BRs to the $c$ and $u$ channels.


\section{FCNC processes in the Higgs boson channel}

When considering FCNC processes where the top quark decays through a Higgs boson both ATLAS~\cite{ATLASH} and CMS~\cite{CMSH} follow a similar strategy. The searches are designed to identify a pair of top quarks in which one of them decays through the usual SM channel while the other one decays through the FCNC decay. Both collaborations present results for data collected at $\sqrt{s}=8$~TeV with an integrated luminosity of $20.3~\rm{fb}^{-1}$ and $19.7~\rm{fb}^{-1}$ for ATLAS and CMS respectively. Different topologies are studied, being sensitive to three different decay channels of the Higgs boson: topologies with high multiplicity of jets and $b$-jets (sensitive to the $H\rightarrow b\bar b$ decay channel), multi-leptonic topologies (sensitive to $H\rightarrow W^+W^-$ and $H\rightarrow \tau^+\tau^-$) and the di-photon channel (sensitive to $H\rightarrow \gamma\gamma$). The three channels are then combined to get the final results.

For the $H\rightarrow b\bar b$ channel in ATLAS, two probability density functions (PDF) are defined based on kinematical information from the different reconstructed objects which are combined into a discriminant variable ($D(\mathbf{x})$) which allows to differentiate signal from background. In CMS two different multi-variate analyses are used. First, a BDT is used to select the best combination of objects to reconstruct the top quark decay and once the event has been reconstructed, a neural-network is used to define a template fit allowing the extraction of signal from background. For the multi-leptonic channel, ATLAS reinterprets the results shown in Ref.~\cite{ATLASHML} with small modifications to ensure orthogonality between the different channels. The analysis is divided in different categories based on the multiplicity of leptons, jets and taus. In CMS the analysis is focused in the $H\rightarrow W^+W^-$ channel, defining a tri-lepton and a same-sign di-lepton channel which are later combined.
The di-photon channel is the cleanest of all three due to the pair of high-$\pt$ photons in the final state. A similar strategy is followed by ATLAS and CMS, defining two different channels based on the decay of the $W$ boson from the top quark decay: a hadronic channel with higher statistics in which the $m_{\gamma\gamma}$  spectrum is used as discriminant variable and a multi-lepton channel in which a counting experiment is done.

Observed upper limits on the BRs for $t\rightarrow Hq$ are set at a 95\% C.L. The combination result for ATLAS exclude BRs above 0.45\% (0.46\%) for the $t\rightarrow Hc$ ($t\rightarrow Hu$) channel. In the CMS combination BRs are excluded above 0.40\% (0.55\%) for the $t\rightarrow Hc$ ($t\rightarrow Hu$) channel. Observed upper limits are also set at 95\% C.L. in the coupling of the $tHq$ vertex: ATLAS excludes $|\lambda_{tHu}| > 0.13$ and $|\lambda_{tHc}| > 0.13$ (Figure~\ref{fig}~(top-right) shows the upper limits as a function of both couplings) and CMS excludes $|\lambda_{tu}^H|^2 > 9.8\times 10^{-3}$ and $|\lambda_{tc}^H|^2 > 6.9\times 10^{-3}$.


\section{FCNC processes in the $\gamma$  channel}

The CMS Collaboration presented a search for the single production of a top quark in association with a photon via FCNC~\cite{CMSGAMMA}. Data collected at $\sqrt{s} = 8$~TeV with an integrated luminosity of $19.8~\rm{fb}^{-1}$ were analysed. The analysis is focused on the leptonic decay of the top quark when the $W$ boson decays to $\mu\nu$ given that this channel presents a cleaner signature (no electrons are allowed). The topology is characterised by the presence of a high-$\pt$ isolated photon, due to the recoil with the top quark, and at least one jet with at most one of the jets being a $b$-tagged jet. Events with more $b$-tagged jets are rejected in order to reduce the $t\bar t+\gamma$ background. Two different multi-variate techniques are used: first, a neural network is trained to identify $W\gamma$+jets and $W$+jets events. Based on the output of this neural network a template fit is done in data to extract the contribution from both background processes. After that, two BDTs are used to discriminate signal from background, one for the $t\gamma u$ vertex and another for $t\gamma c$. The output of this BDT is used as discriminant variable and upper limits in the BRs and couplings ($\kappa_{tu\gamma}$ and $\kappa_{tc\gamma}$) are set at a 95\% C.L. The BRs above $1.3\times 10^{-4}$ ($1.7\times 10^{-3}$) are excluded for the $t\rightarrow u\gamma$ ($t\rightarrow c\gamma$) process. The upper limits set for the couplings are $\kappa_{tu\gamma} < 0.025$ and $\kappa_{tc\gamma} < 0.091$.

\section{FCNC processes in the gluon channel}

When considering FCNC processes in the gluon channel searches are designed to to target direct top production ($u/cg\rightarrow t$) studying FCNC in the production instead of the decay of the top quark.

The ATLAS Collaboration~\cite{ATLASGLU} analysed data collected at $\sqrt{s}=8$~TeV with an integrated luminosity of $20.3~\rm{fb}^{-1}$. In this analysis the top quark decays leptonically and multi-jet background is reduced by requiring the $\pt$ of the lepton to satisfy

\beq
\pt^{\ell} > 90\rm{ GeV}\left(1-\frac{\pi-|\Delta\phi(\ell,jet)|}{\pi-2}\right)
\eeq{ptl} 

\begin{figure}[!t]
\centering
\includegraphics[width=.48\linewidth]{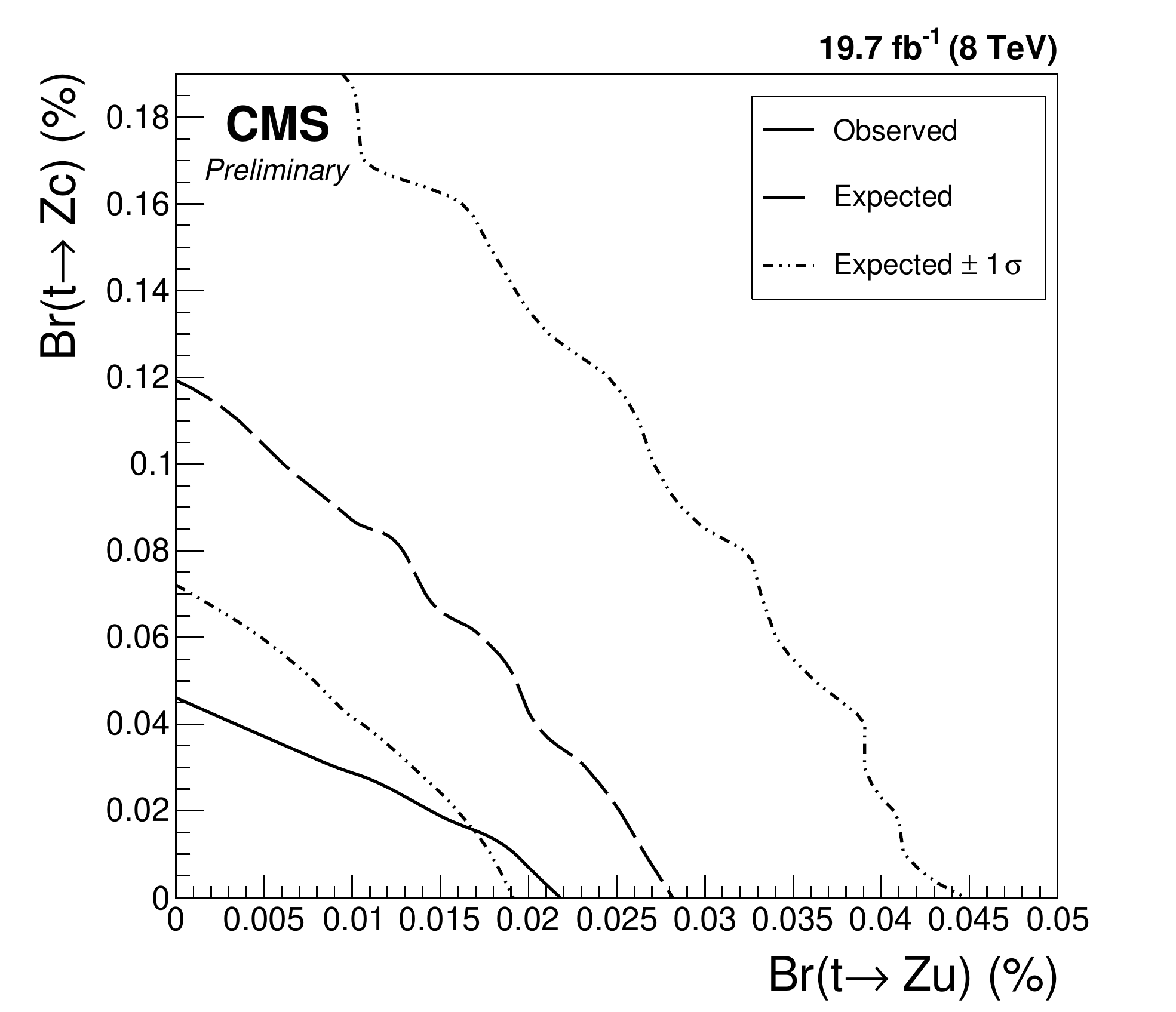}
\includegraphics[width=.48\linewidth]{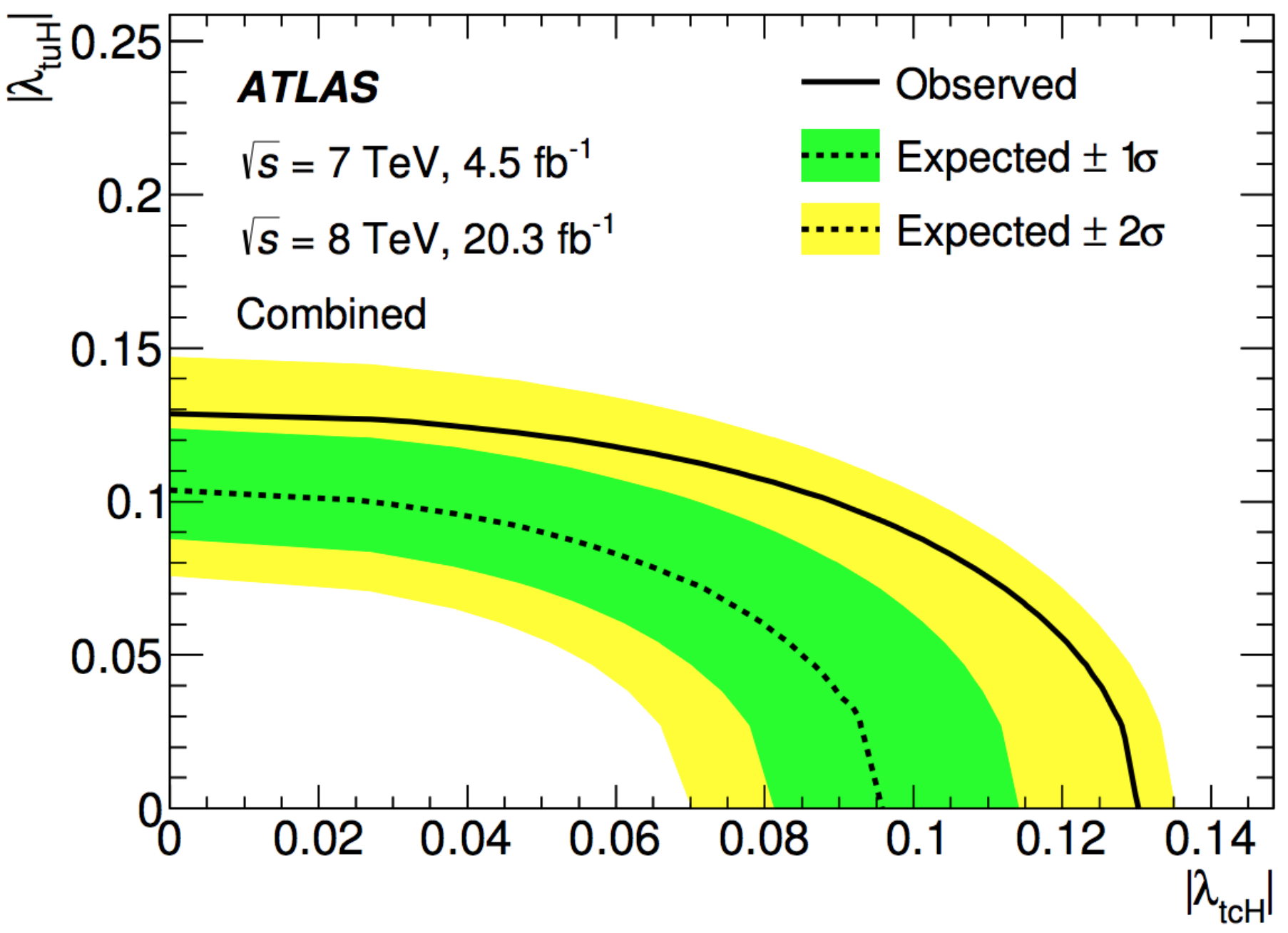}
  \includegraphics[width=.48\textwidth]{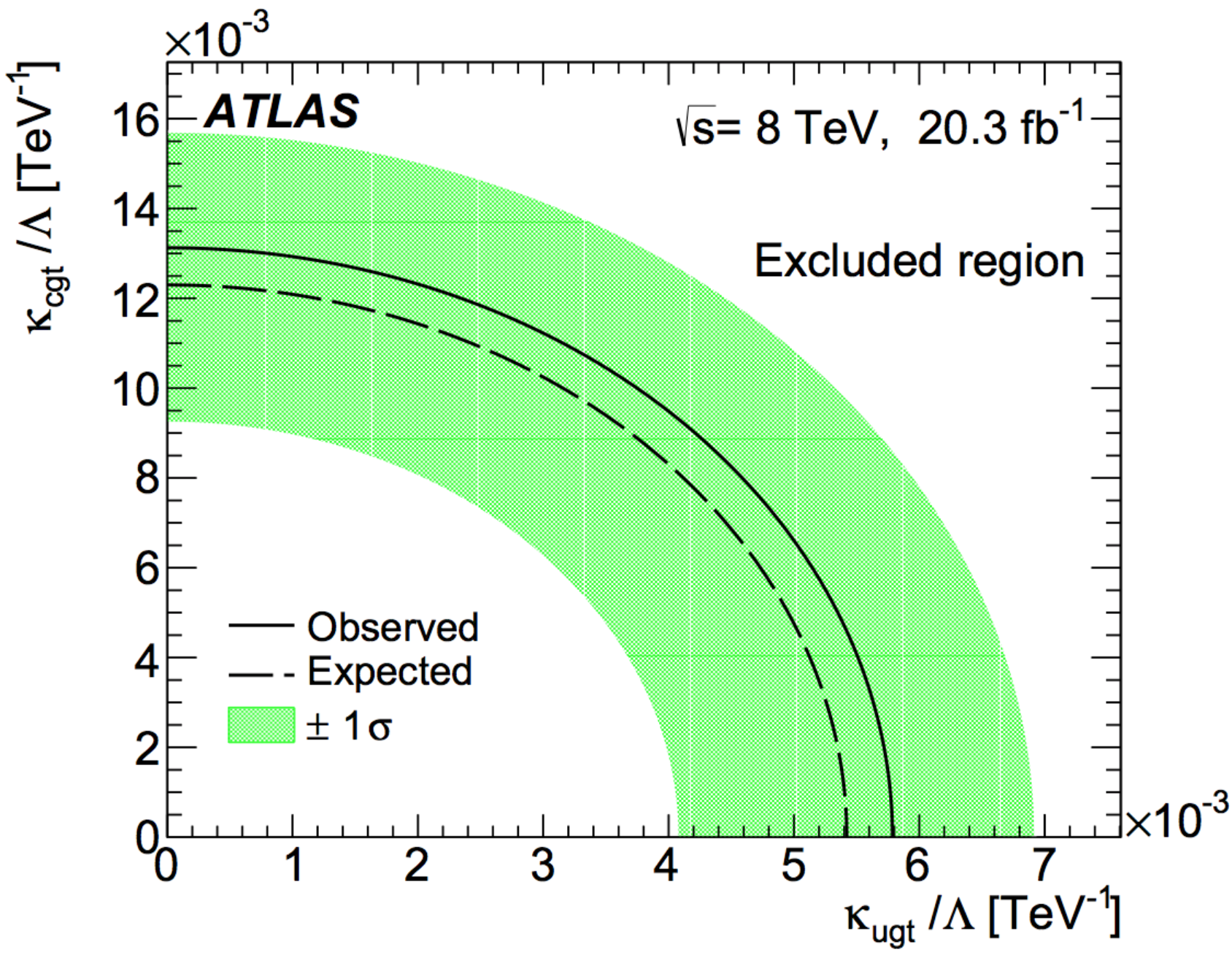}
  \includegraphics[width=.48\linewidth]{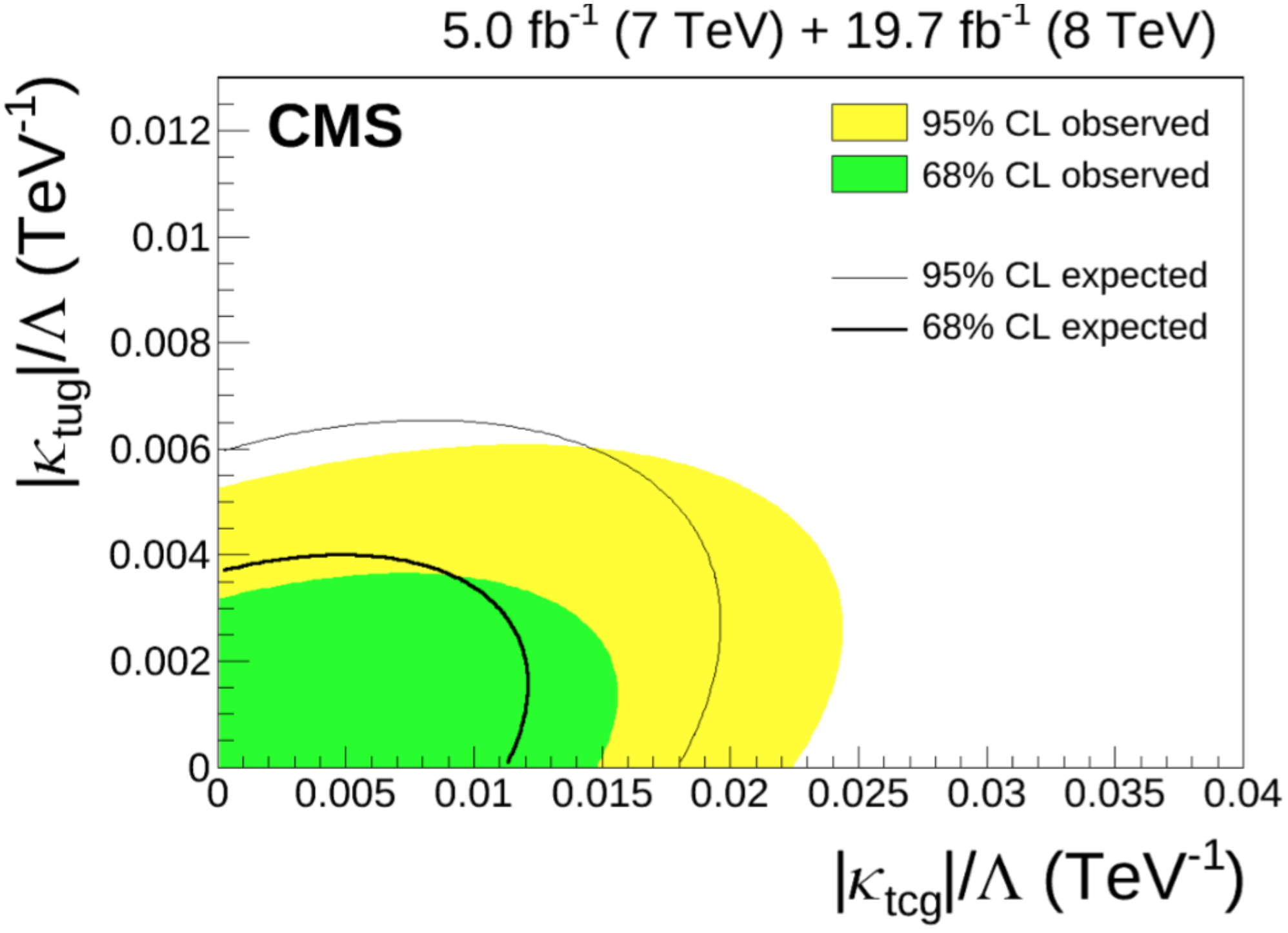}
\caption[Limits exclusions]{Excluded regions as a function of the FCNC couplings for different channels: $Z$ channel (top-left)~\cite{CMSZ}, Higgs boson channel (top-right)~\cite{ATLASH} and gluon channel from ATLAS (bottom-left)~\cite{ATLASGLU} and CMS (bottom-right)~\cite{CMSG}.}
\label{fig}
\end{figure}

In order to differentiate signal from background a neural network is defined with several input distributions with discriminant power (e.g. transverse mass of the top quark and the lepton $\pt$). The shape of the output of the neural network is used to set limit which exclude, with a 95\% C.L. BRs above $4\times 10^{-5}$ ($2\times 10^{-4}$) for $t\rightarrow ug$ ($t\rightarrow cg$). The observed upper limits for the couplings ($\kappa_{ugt}$ and $\kappa_{cgt}$) at 95\% C.L. are set at $\kappa_{ugt}/\Lambda > 5.8\times 10^{-3}\rm{ TeV}^{-1}$ and $\kappa_{cgt}/\Lambda > 1.3\times 10^{-2}\rm{ TeV}^{-1}$. Figure~\ref{fig}~(bottom-left) shows the excluded region as a function of the couplings.

The CMS Collaboration analysed data collected at 7 and 8~TeV with an integrated luminosity of $5~\rm{fb}^{-1}$ and $19.7~\rm{fb}^{-1}$ respectively~\cite{CMSG}. The analysis is focused on the leptonic decay of the quark as well requiring one isolated muon rejecting events with extra leptons. The event should have 2 or 3 jets with at least one of them tagged as a $b$-jet. In order to suppress the multi-jet background a neural is trained recognise it and a cut is applied on the output of the neural network highly reducing the multi-jet contamination. After that, two neural networks are trained to differentiate signal from background, one for the $tcg$ vertex and another for the $tug$ vertex. The output of both neural networks is used as discriminant variables and observed upper limits are set on the BRs and couplings ($\bar\kappa_{ugt}$ and $\bar\kappa_{cgt}$) for the FCNC process, excluding, at a 95\% C.L., BRs above $2\times 10^{-5}$ ($4.1\times 10^{-4}$) for $tug$ ($tcg$) vertices and couplings with values $\bar\kappa_{ugt}/\Lambda > 4.1\times 10^{-2}\rm{ TeV}^{-1}$ and $\kappa_{cgt}/\Lambda > 1.8\times 10^{-2}\rm{ TeV}^{-1}$. Figure~\ref{fig}~(bottom-right) shows the excluded region as a function of the couplings.

\begin{thebibliography}{99}


\bibitem{ATLAS} 
  ATLAS Collaboration,
  ``The ATLAS Experiment at the CERN Large Hadron Collider'',
  JINST 3, S08003 (2008).
  
\bibitem{CMS} 
  CMS Collaboration,
  ``The CMS experiment at the CERN LHC'',
  JINST {\bf 3}, S08004 (2008).

\bibitem{ATLASZ} 
  ATLAS Collaboration,
  ``Search for flavour-changing neutral current top-quark decays to $qZ$ in $pp$ collision data collected with the ATLAS detector at $\sqrt s =8$ ~TeV'',
  Eur.\ Phys.\ J.\ C 76, no. 1, 12 (2016)
  [arXiv:1508.05796 [hep-ex]].

\bibitem{CMSZ} 
  CMS Collaboration,
  ``Search for associated production of a Z boson with a single top quark and for tZ flavour-changing interactions in pp collisions at $\sqrt{s} = 8$~TeV'',
  CMS-PAS-TOP-12-039.

\bibitem{ATLASH} 
  ATLAS Collaboration,
  ``Search for flavour-changing neutral current top quark decays $t\to Hq$ in $pp$ collisions at $\sqrt{s}=8$~TeV with the ATLAS detector'',
  JHEP 1512, 061 (2015)
  [arXiv:1509.06047 [hep-ex]].
  
  \bibitem{CMSH} 
  CMS Collaboration,
  ``Search for top quark decays via Higgs-boson-mediated flavor-changing neutral currents in pp collisions at $\sqrt{s} = 8$~TeV'',
  CMS-TOP-13-017,
  arXiv:1610.04857 [hep-ex].
  
  

\bibitem{ATLASHML} 
  ATLAS Collaboration,
  ``Search for the associated production of the Higgs boson with a top quark pair in multilepton final states with the ATLAS detector'',
  Phys.\ Lett.\ B {749}, 519 (2015)
  [arXiv:1506.05988 [hep-ex]].
  
\bibitem{CMSGAMMA} 
  CMS Collaboration,
  ``Search for anomalous single top quark production in association with a photon in pp collisions at $ \sqrt{s}=8 $ TeV'',
  JHEP {1604}, 035 (2016)
  [arXiv:1511.03951 [hep-ex]].
\bibitem{ATLASGLU} 
  ATLAS Collaboration,
  ``Search for single top-quark production via flavour-changing neutral currents at 8 TeV with the ATLAS detector'',
  Eur.\ Phys.\ J.\ C 76, no. 2, 55 (2016)
  [arXiv:1509.00294 [hep-ex]].
  
  \bibitem{CMSG} 
  CMS Collaboration,
  ``Search for anomalous $Wtb$ couplings and flavour-changing neutral currents in t-channel single top quark production in $pp$ collisions at $\sqrt{s} = 7$ and 8~TeV'',
  	CMS-TOP-14-007,
  arXiv:1610.03545 [hep-ex].
  
  
  
\end{thebibliography}
\end{document}